\documentclass[aps,prl,twocolumn,superscriptaddress,nopacs]{revtex4}
\usepackage{graphicx}
\usepackage{dcolumn}

\usepackage{bm}
\usepackage{color}
\usepackage[colorlinks]{hyperref}
\input{epsf}
\usepackage{verbatim}
\begin{document}

%\preprint{}

\title{Phase study of oscillatory resistances in microwave-irradiated- and dark- GaAs/AlGaAs devices:
Indications of a new class of integral quantum Hall effect}

\author{R. G. Mani}
\affiliation{Department of Physics and Astronomy, Georgia State
University,   29 Peachtree Center Avenue, Atlanta, GA 30303 U.S.A.
}

\author{W. B. Johnson}
 \affiliation{Laboratory for Physical Sciences, University of
 Maryland, College Park, MD 20740 U.S.A.}

\author{V. Umansky}
\affiliation{Braun Center for Submicron Research, Weizmann
Institute, Rehovot 76100, Israel}

\author{V. Narayanamurti}
\affiliation{SEAS, Harvard University, 29 Oxford Street,
Cambridge, MA 02138 U.S.A.}

\author{K. Ploog}
\affiliation{Paul-Drude-Institut f\"{u}r Festk\"{o}rperelektronik,
Hausvogteiplatz 5-7, 10117 Berlin, Germany}

%\author{}
%\thanks{}
%\affiliation{}
\date{\today. Jour. Ref: Phys. Rev. B 79, 205320 (2009)}
\begin{abstract}
We report the experimental results from a dark study and a
photo-excited study of the high mobility GaAs/AlGaAs system at
large filling factors, $\nu$. At large-$\nu$, the dark study
indicates several distinct phase relations ("Type-1", "Type-2",
and "Type-3") between the oscillatory diagonal- and Hall-
resistances, as the canonical Integral Quantum Hall Effect (IQHE)
is manifested in the "Type-1" case of approximately orthogonal
diagonal- and Hall resistance- oscillations. Surprisingly, the
investigation indicates quantum Hall plateaus also in the "Type-3"
case characterized by approximately "anti-phase" Hall- and
diagonal- resistance oscillations, suggesting a new class of IQHE.

Transport studies under microwave photo-excitation exhibit
radiation-induced magneto-resistance oscillations in both the
diagonal, $R_{xx}$, and off-diagonal, $R_{xy}$, resistances.
Further, when the radiation-induced magneto-resistance
oscillations extend into the quantum Hall regime, there occurs a
radiation-induced non-monotonic variation in the amplitude of
Shubnikov-de Haas (SdH) oscillations in $R_{xx}$ \textit{vs}. B,
and a non-monotonic variation in the width of the quantum Hall
plateaus in $R_{xy}$. The latter effect leads into the vanishing
of IQHE at the minima of the radiation-induced $R_{xx}$
oscillations with increased photo-excitation. We reason that the
mechanism which is responsible for producing the non-monotonic
variation in the amplitude of SdH oscillations in $R_{xx}$ under
photo-excitation is also responsible for eliminating, under
photo-excitation, the novel "Type-3" IQHE in the high mobility
specimen.
\end{abstract}
%\pacs{73.40.-c,73.43.Qt, 73.43.-f, 73.21.-b}
%\pacs{}
%\keywords{}
\maketitle

\section{introduction}
A 2-dimensional electron system (2DES) at high magnetic fields,
$B$, and low temperatures, $T$, exhibits the integral quantum Hall
effect (IQHE), which is characterized by plateaus in the Hall
resistance $R_{xy}$ \textit{vs}. $B$, at $R_{xy} = h/ie^{2}$, with
$i$ = 1,2,3,... and concurrent vanishing diagonal resistance
$R_{xx}$ as $T \rightarrow$ 0 K, in the vicinity of integral
filling factors of Landau levels, i.e., $\nu \approx
i$.\cite{1,2,3} With the increase of the electron mobility, $\mu$,
at a given electron density, $n$, and low $T$, IQHE plateaus
typically become narrower as fractional quantum Hall effects
(FQHE) appear in the vicinity of $\nu \approx p/q$, at $R_{xy} =
h/[(p/q)e^{2}]$, where $p/q$ denotes mostly odd-denominator
rational fractions.\cite{2,3} Experimental studies of the highest
mobility specimens have typically focused upon FQHE and other
novel phases.\cite{2,3,4,5} Meanwhile, the possibility of new
variations of IQHE that might appear with the canonical effect in
the reduced-disorder specimen, especially at large-$\nu$, has been
largely unanticipated. In the first dark-study part of this paper,
we show that three distinct phase relationships can occur between
the oscillatory diagonal- and Hall- resistances in the
high-mobility dark specimen at $\nu > 5$, and that IQHE can be
manifested in two of these variations. The results therefore
suggest one new class of IQHE, as they provide insight into the
origin of oscillatory variations in the Hall effect, and their
evolution into Hall plateaus, in the low-$B$ large-$\nu$ regime of
the radiation-induced zero-resistance states in the photoexcited
high mobility 2DES.\cite{6, 7, 8, 9, 10, 11, 12, 13, 14, 15, 16,
17, 18, 19, 20, 21, 22, 23, 24, 25, 26, 27, 28, 29, 30, 31, 32,
33, 34, 35, 36, 37, 38, 39, 40, 41, 42, 43, 44, 45, 46, 47, 48,
49}
\begin{figure}[t]
%h=here, t=top, b=bottom, p=separate figure page
\begin{center}
\leavevmode \epsfxsize=3.25in
 \epsfbox {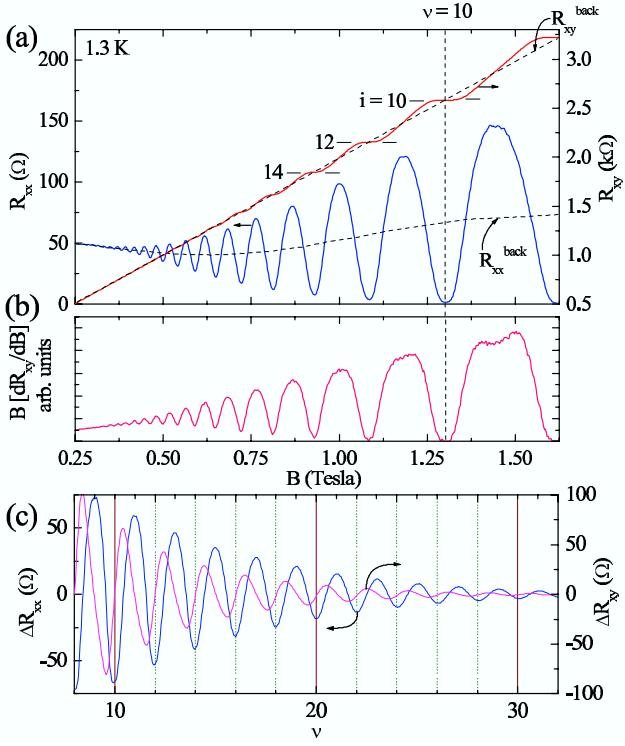}
\end{center}
\caption{Canonical IQHE and "Type-1" oscillations: (a) Hall
plateaus at $R_{xy} = h/ie^{2}$, i.e., IQHE's, coincide with
minima in the diagonal resistance $R_{xx}$ in a GaAs/AlGaAs Hall
bar device. (b) A comparison of $B[dR_{xy}/dB]$ shown here in Fig.
1(b) with $R_{xx}$ in Fig. 1(a) suggests that $R_{xx} \propto
B[dRxy/dB]$, as per the resistivity rule (ref. 50). (c) A
canonical quantum Hall system at large-$\nu$ also exhibits an
approximately $\pi/2$ phase shift between the oscillatory parts of
$R_{xx}$ and $R_{xy}$. } \label{mani01fig}
\end{figure}

Transport studies of photo-excited two-dimensional electron
systems have become a topic of interest following the discovery of
novel radiation-induced zero-resistance states in the GaAs/AlGaAs
system at high filling factors.\cite{6,16} The characteristic
field $B_{f}$ for these zero-resistance states and associated
magneto-resistance oscillations is a linear function of the
radiation frequency, $f$, i.e., $B_{f} = 2\pi f m^{*}/e$, and,
therefore, an increase in $f$ could be expected to bring about an
overlap of the radiation-induced zero-resistance states and the
quantum Hall effect. In this case, a topic of interest is the
interplay between the radiation-induced phenomena and quantum Hall
effect. Hence, in the second part of the paper, we follow up the
above-mentioned dark-study of oscillatory resistances and quantum
Hall effect at large filling factors, by examining also the
influence of microwave photo-excitation. Here, the experimental
results show that vanishing resistance induced by photo-excitation
helps to replace the quantum Hall effect by an ordinary Hall
effect over broad magnetic field intervals in the vicinity of the
radiation-induced oscillatory $R_{xx}$ minima. The results also
identify a strong correlation between the vanishing of SdH
oscillations in $R_{xx}$ and the narrowing of Hall plateaus in
$R_{xy}$ under photo-excitation. Comparative plots of the
oscillatory diagonal and off-diagonal resistances that help to
establish this correlation also serve to confirm the observations
made on the dark specimen, namely, that IQHE sometimes goes
together with "Type-3" resistance oscillations in the high
mobility specimen at large $\nu$.

\begin{figure}[t]
%h=here, t=top, b=bottom, p=separate figure page
\begin{center}
\leavevmode \epsfxsize=3.25in
 \epsfbox {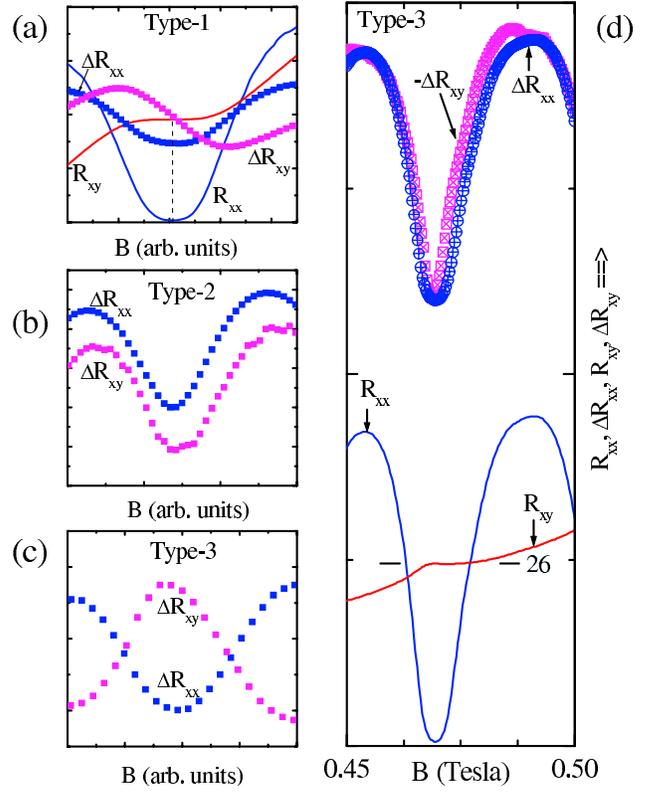}
\end{center}
\caption{In this figure, panels (a)-(c) exhibit, using
representative data, the three types of magnetoresistance
oscillations that are examined in this study
 while panel (d) illustrates an
experimental observation of novel IQHE observed in the case of
"Type-3" oscillations. (a) This panel shows the
approximate-orthogonality observed between the $\Delta R_{xx}$ and
$\Delta R_{xy}$ oscillations at large filling factors in the usual
IQHE, see Fig. 1(c). (b) This panel illustrates the "in-phase"
("Type-2") oscillations in $\Delta R_{xx}$ and $\Delta R_{xy}$ at
large filling factors, which have been observed in this study. (c)
This panel presents the approximately $\pi$ phase shift between
$\Delta R_{xx}$ and $\Delta R_{xy}$ oscillations that
characterizes Type-3 oscillations, also observed at large $\nu$ in
this study. (d) At the top, this panel exhibits "Type-3"
oscillations at large filling factors. The bottom part of the
panel (d) illustrates the associated IQHE.} \label{mani01fig}
\end{figure}
\section{experiment}
Simultaneous low-frequency ac lock-in based electrical
measurements of $R_{xx}$ and $R_{xy}$ were carried out on
GaAs/AlGaAs single hetero-junctions at $T
> 0.45 K$, with matched lock-in time constants, and sufficiently
slow $B$-field sweep rates.  The $B$-field was calibrated by ESR
of DPPH.\cite{12, 29}
\begin{figure}[t]
%h=here, t=top, b=bottom, p=separate figure page
\begin{center}
\leavevmode \epsfxsize=3.0in
 \epsfbox {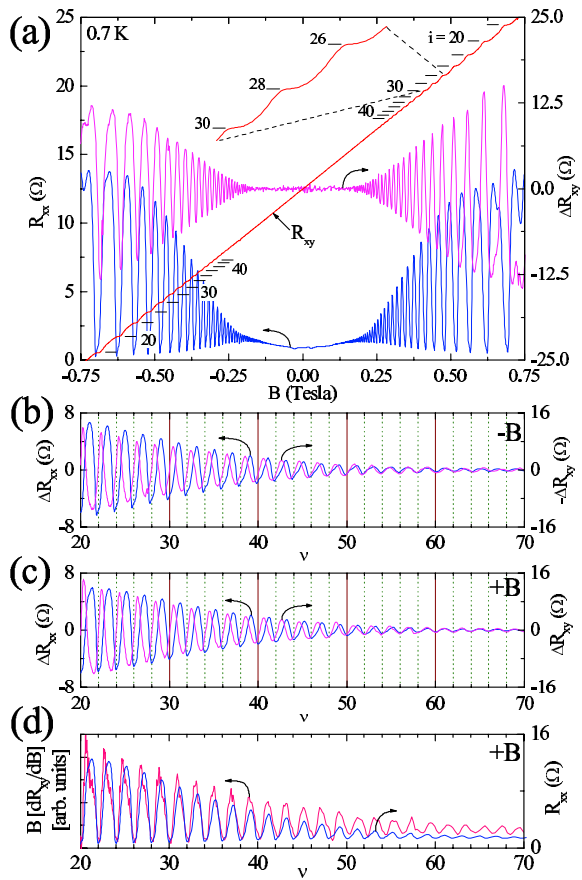}
\end{center}
\caption{"Type-1" to "Type-2" crossover at large-$\nu$: (a)
$R_{xx}$, $R_{xy}$, and the oscillatory Hall resistance, $\Delta
R_{xy}$, are shown over low magnetic fields, $B$, for a high
mobility GaAs/AlGaAs specimen. Here, the oscillatory Hall
resistance is anti-symmetric under field reversal, i.e., $\Delta
R_{xy}(-B) = -\Delta R_{xy}(+B)$, as expected. (b) The oscillatory
diagonal resistance ($\Delta R_{xx}$) and $\Delta R_{xy}$ have
been plotted \textit{vs}. $\nu$ to compare their relative phases
for $-B$. (c) As above for $+B$. For $20 \leq \nu < 46$, $\Delta
R_{xx}$ and $\Delta R_{xy}$ are approximately orthogonal as in
Fig. 1(c). For $56 \leq \nu \leq 70$, $\Delta R_{xx}$ and $\Delta
R_{xy}$ are approximately in-phase, unlike at $\nu < 46$. Note
that the right ordinates in Fig. 3(c) and Fig. 3(b) show $+\Delta
R_{xy}$ and $-\Delta R_{xy}$, respectively, in order to account
for the antisymmetry in $\Delta R_{xy}$ under $B$-reversal. (d)
$B(dR_{xy}/dB)$ and $R_{xx}$ are plotted \textit{vs}. $\nu$. For
$\nu < 46$, the two quantities are similar and in-phase, while a
phase difference develops at higher $\nu$.} \label{mani02fig}
\end{figure}

Shubnikov-de Haas oscillations and associated IQHE became weaker,
as usual, at higher $T$, and few oscillations or Hall plateaus
were evident for $\nu > 20$ at $T > 1.7 K$  in the dark study.
Thus, we focused upon $0.45 < T < 1.7 K$, where $T$-induced
changes in the phase relations were not discerned and, further,
beats were not observed in the SdH oscillations. The observed
phase differences were verified not to be experimental artifacts
originating from the choice of experimental parameters such as the
B-sweep rate, the data acquisition rate, lock-in integration time,
and other typical variables. The observed phase relations also did
not show an obvious dependence on the sample geometry, or type
(Au-Ge/Ni or In) of contacts. The reported phase relation between
the oscillatory Hall- and diagonal- resistances could often be
identified by eye. Yet, we have utilized background subtraction
here for the sake of presentation, mainly to realize overlays in
the figures, for phase comparison. An explicit example of the
background subtraction procedure is illustrated in Appendix A.
Finally, although the mobility $\mu$ has been provided, $\mu$
alone seems not to be sufficient for classifying the observed
phenomena in high-$\mu$ specimens. Here, the high mobility
condition was realized by brief illumination with a red LED.
\begin{figure}[t]
%h=here, t=top, b=bottom, p=separate figure page
\begin{center}
\leavevmode \epsfxsize=3.0in
 \epsfbox {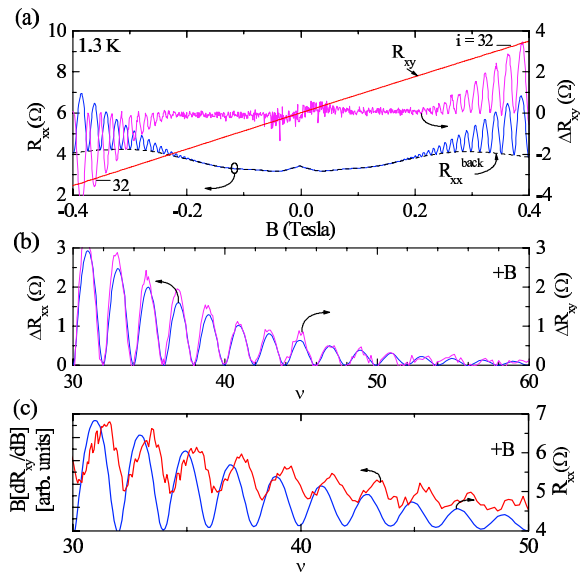}
\end{center}
\caption{"Type-2" oscillations at large-$\nu$: (a) $R_{xx}$,
$R_{xy}$, and $\Delta R_{xy}$ have been shown for a Hall bar
specimen with $n = 2.9 \times 10^{11} cm^{-2}$ and $\mu = 6 \times
10^{6} cm^{2}/V s$, which exhibits in-phase $R_{xx}$ and $\Delta
R_{xy}$ oscillations. Note the absence of discernable Hall
plateaus in $R_{xy}$, when the oscillatory resistances are
in-phase. (b) Here, the amplitudes of $\Delta R_{xx}$ and $\Delta
R_{xy}$ show similar $\nu$ -variation. (c) $B(dR_{xy}/dB)$ and
$R_{xx}$ are plotted \textit{vs}. $\nu$. A comparison of the two
traces suggests a $90^{0}$ phase shift between the two
quantities.} \label{mani02fig}
\end{figure}

\section{Summary of the other observable phase relations at large-$\nu$}

\begin{figure}[t]
%h=here, t=top, b=bottom, p=separate figure page
\begin{center}
\leavevmode \epsfxsize=2.5in
 \epsfbox {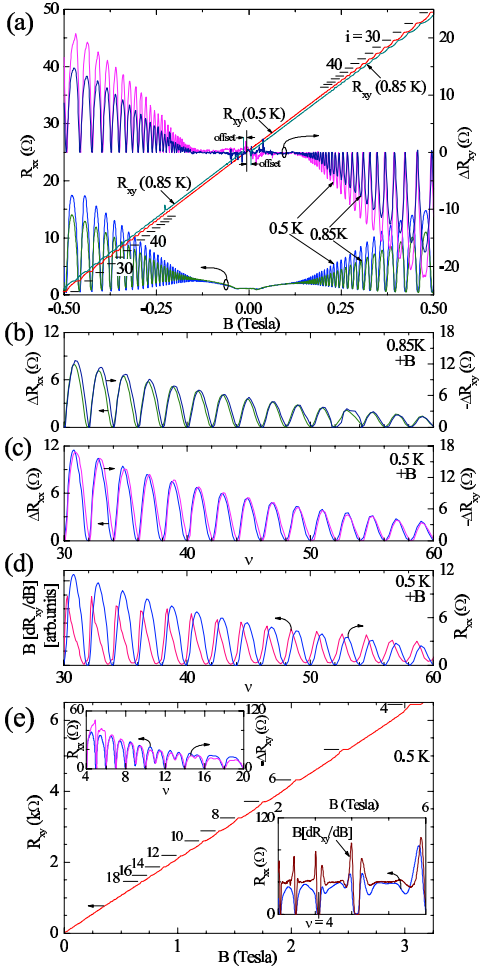}
\end{center}
\caption{"Type-3" oscillations at large-$\nu$, and "Type-3" to
"Type-1" crossover at large-$B$: (a) Data for a square shape
GaAs/AlGaAs specimen, where the magnitude of $R_{xy}$ is reduced
at the $R_{xx}$ oscillation maxima. Here, the $+B$ and the $-B$
portions of the $R_{xy} (0.85 K)$ curve have been offset in
opposite directions along the abscissa with respect to the $R_{xy}
(0.5 K)$ curve, for the sake of presentation. Note the
well-developed plateaus in $R_{xy}$. (b) At $T= 0.85 K$, a plot of
$\Delta R_{xx}$ and $-\Delta R_{xy}$ confirms similarity and a
Type-3 phase relation. (c) Same as (b) but at $T= 0.5 K$. (d)
$B(dR_{xy}/dB)$ and $R_{xx}$ are plotted \textit{vs}. $\nu$. The
plot shows incommensurability between the two quantities. (e) The
main panel shows the Hall resistance \textit{vs}. $B$, with
relatively narrow Hall plateaus, down to around filling factor
$\nu = 4$. Left Inset: As in Fig. 5(b) and 5(c) above, $-\Delta
R_{xy}$ follows $R_{xx}$, down to nearly $\nu = 5$. Right Inset:
For $\nu \leq 4$, a better correspondence develops between
$R_{xx}$ and $B[dR_{xy}/dB]$. } \label{mani03fig}\end{figure}

The second part of the study, i.e., the photo-excited study,\cite
{6} followed the same methods as the dark study  with the
difference that microwaves introduced via a rectangular waveguide
served to irradiate the specimen. Here, the radiation intensity
was adjusted externally as desired.
\section{Part 1 - The dark study}
\section{Background: "Type-1" phase relation in the canonical IQHE at large-$\nu$}
To review the basics, figure 1(a) exhibits measurements from a
typical low mobility Hall bar specimen with $n = 3.2 \times
10^{11} cm^{-2}$ and $\mu = 0.4 \times 10^{6} cm^{2}/Vs$. Here, as
is usual with IQHE, large amplitude Shubnikov-de Haas (SdH)
oscillations in $R_{xx}$ lead into zero-resistance states with
increasing $B$, as $R_{xy}$ exhibits plateaus at $R_{xy} =
h/ie^{2}$ for $\nu \approx i$, with $i$ = 2,4,6,... This canonical
low-mobility IQHE system is known to follow a
resistivity/resistance rule,\cite{50,51} at a each $T$,\cite{52}
whereby $R_{xx} \propto B[dR_{xy}/dB]$ and $dR_{xy}/dB$ is the
$B$-field derivative of $R_{xy}$.\cite{50} In order to check the
validity  of this rule, Fig. 1(b) exhibits $B[dR_{xy} /dB]$, which
is then to be compared with the $R_{xx}$ shown in Fig. 1(a). Such
a comparison suggests general consistency between the observed
results and the suggested rule.\cite{50,51,52}

For the sake of further analysis, Fig. 1(c) shows the oscillatory
part of the diagonal ($\Delta R_{xx}$) and the off-diagonal Hall
($\Delta R_{xy}$) resistances \textit{vs}. $\nu$. Here, $\Delta
R_{xy} = R_{xy} - R_{xy}^{back}$ and $\Delta R_{xx} = R_{xx} -
R_{xx}^{back}$, as $R_{xy}^{back}$ and $R_{xx}^{back}$ are the
background resistances shown in Fig. 1(a). Note that, sometimes,
we shall also refer to this oscillatory $\Delta R_{xy}$ as "SdH
oscillations", for lack of a better term, simply to avoid
confusing it with the radiation-induced oscillations in $R_{xy}$.
The reader should bear in mind that these "SdH oscillations" in
$R_{xy}$ are sometimes just another manifestation of IQHE. As
evident from Fig. 1(c), the quantum Hall characteristics of Fig.
1(a) yield approximately orthogonal oscillations in $\Delta
R_{xx}$ and $\Delta R_{xy}$ such that $\Delta R_{xx} \approx
-cos(2\pi[\nu/2])$ and $\Delta R_{xy} \approx sin(2\pi[\nu/2])$.
This phase description becomes more appropriate at higher filling
factors as the harmonic content in the oscillations is reduced,
and the oscillations attain the appearance of exponentially damped
sine/cosine waves. We denote the quantum Hall features of Fig.
1(a)-(c) as "Type-1" characteristics, and present the essentials
in Fig. 2(a).

This study reports on other observable phase relations in the high
mobility 2DES. We find, for instance, a "Type-2" phase relation,
where $|R_{xy}|$ is enhanced (suppressed) at the $R_{xx}$
oscillation peaks (valleys) and the $\Delta R_{xy}$ oscillations
are in-phase with the $R_{xx}$ or $\Delta R_{xx}$ oscillations, as
shown in Fig. 2(b). There also occurs the more remarkable "Type-3"
case where $|R_{xy}|$ is enhanced (suppressed) at the $R_{xx}$ SdH
oscillation minima (maxima) and the $\Delta R_{xy}$ oscillations
are phase-shifted by approximately "$\pi$" with respect to the
$R_{xx}$ or $\Delta R_{xx}$ oscillations, as shown in Fig. 2(c).
Both Type-2 and Type-3 oscillations show variance from the
resistivity/resistance rule for quantum Hall systems.\cite{50}
Here, we survey these experimentally observed phase relations and
related crossovers in the high mobility 2DES, and then focus on
the "Type-3" case, which also brings with it, remarkably, a new
class of IQHE that is illustrated in Fig. 2(d).

\begin{figure}[t]
%h=here, t=top, b=bottom, p=separate figure page
\begin{center}
\leavevmode \epsfxsize=3.25in \epsfbox {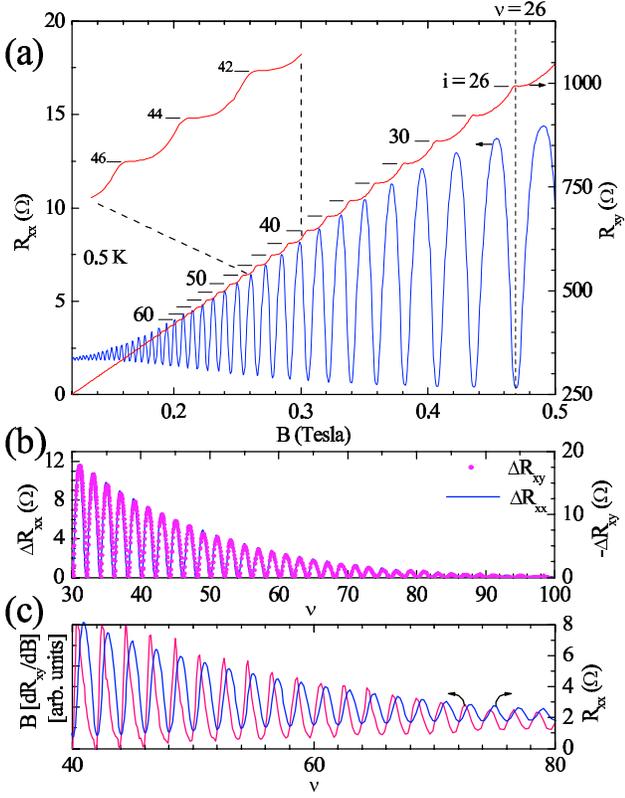}
\end{center}
\caption {IQHE with "Type-3" oscillations: (a) $R_{xx}$ and
$R_{xy}$ are exhibited for a high mobility GaAs/AlGaAs specimen
that shows deep $R_{xx}$ minima and even-integral Hall plateaus at
$R_{xy} = h/ie^{2}$. (b) Here, $-\Delta R_{xy}$ follows $\Delta
R_{xx}$, indicative of a $\pi$ phaseshift and a "Type-3"
relationship, unlike the canonical quantum Hall situation, see
Fig. 1(c). (c) $B(dR_{xy}/dB)$ and $R_{xx}$ are plotted
\textit{vs}. $\nu$ to compare with the resistivity/resistance
rule. Here, the lineshapes look dissimilar, and there is a phase
shift between $B(dR_{xy}/dB)$ and $R_{xx}$.} \label{mani05fig}
\end{figure}
\section{Results}
\section{"Type-1" - "Type-2" crossover and "Type-2" phase relation at large-$\nu$}
For a high mobility specimen with $n = 3 \times 10^{11} cm^{-2}$
and $\mu = 1.1 \times 10^{7} cm^{2}/V s$ that shows IQHE up to $i
\approx 40$, Fig. 3(a) illustrates $R_{xx}$, $R_{xy}$, and $\Delta
R_{xy}$ for both $B$-directions, using the convention $R_{xy} > 0$
for $B > 0$. Figures 3(b) and (c) confirm similar behavior for
both B-directions once the anti-symmetry in $\Delta R_{xy}$ under
$B$-reversal is taken into account. Fig. 3(c) indicates that from
$20 \leq \nu < 46$, $\Delta R_{xy}$ oscillations are approximately
orthogonal to the $\Delta R_{xx}$ oscillations, as in Fig. 1(c).
This feature, plus the manifestation of Hall plateaus in Fig.
3(a), and the consistency with the resistivity/resistance rule
indicated in Fig. 3(d) over this $\nu$-range, confirms that the
IQHE observed here is the canonical effect. A remarkable and
interesting feature in Fig. 3(c) is that, following a smooth
crossover, $\Delta R_{xx}$ and $\Delta R_{xy}$ become in-phase,
i.e., "Type-2", for $\nu \geq $ 56, as a variance with the
resistivity/resistance rule, in the form of a phase shift,
develops in Fig. 3(d).

Figs. 4(a) and 4(b) provide further evidence for in-phase "Type-2"
oscillations in a Hall bar. Here, the Hall oscillations tend to
enhance the magnitude of $R_{xy}$ at the $R_{xx}$ SdH maxima
("Type-2"), even as Hall plateaus are imperceptible in the
$R_{xy}$ curve. Yet, from Fig. 4(a), it is clear that $\Delta
R_{xy}$ is a Hall effect component, and not a misalignment offset
admixture of $R_{xx}$ into $R_{xy}$, since $\Delta R_{xy}$ is
antisymmetric under $B$-reversal.

We have presented these data exhibiting these "Type-2"
oscillations and "Type-1" to "Type-2" crossover mainly for the
sake of completeness; the main focus of this paper are the results
that appear in the following.

\begin{figure}[t]
%h=here, t=top, b=bottom, p=separate figure page
\begin{center}
\leavevmode \epsfxsize=3.25in \epsfbox {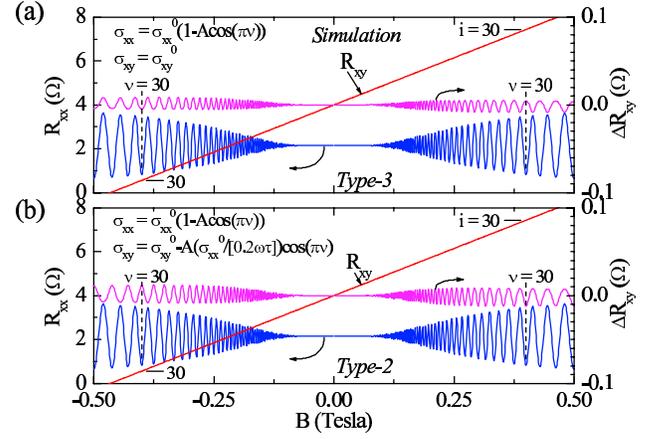}
\end{center}
\caption {Modelling shows the possibility of "Type-3" and "Type-2"
oscillations: (a) Simulations suggest that oscillatory scattering
contributions to the diagonal conductivity alone can produce
Type-3 oscillations via the tensor relation for the resistivities,
where the magnitude of $R_{xy}$ is reduced at the SdH $R_{xx}$
maxima. (b) Simulations of a semiempirical model that includes
both a scattering contribution in $\sigma_{xy}$, and a reduction
in the relaxation time with respect to the transport lifetime,
indicate the possibility also of Type-2 oscillations, with
$|\Delta R_{xy}| << |\Delta R_{xx}|$. A comparison of Fig. 7 (a)
and 7(b) helps to convey the $\tau$-induced Type-3 to Type-2
transformation.} \label{mani05fig}
\end{figure}

\section{"Type-3" phase relation at large-$\nu$ and novel IQHE}

Figure 5 illustrates the third ("Type-3") phase relation in a high
mobility square shape 2DES with $n = 2.9 \times 10^{11} cm^{2}$
and $\mu = 1 \times 10^{7} cm^{2}/Vs$. Although $\mu$ for this
specimen is similar to the one examined in Fig. 3(a)-(d), the
experimental results do look different. Figure 5(a) exhibits data
taken at $T = 0.85 K$ and $T = 0.5 K$. Fig. 5(a) shows that the
main effect of changing $T$ is to modify the amplitude of the
$\Delta R_{xx}$ and $\Delta R_{xy}$ oscillations, so that
oscillatory effects persist to a lower $B$ at the lower $T$. The
data of Fig. 5(a) also show that $\Delta R_{xy}$ tends to reduce
the magnitude of $R_{xy}$ over the B-intervals corresponding to
the $R_{xx}$ peaks, as in Fig. 2(c), the Type-3 case. Meanwhile,
quantum Hall plateaus are easily perceptible in $R_{xy}$, see Fig
5(a). Figures 5(b) and 5(c) demonstrate that for $+B$, for
example, $\Delta R_{xx}$ and $-\Delta R_{xy}$ show nearly the same
lineshape for $30 \le \nu \le 60$, and the phase relation does not
change with $T$. Meanwhile, a comparison of $B[dR_{xy}/dB]$ and
$R_{xx}$, see Fig. 5(d), suggests a variance with the
resistivity/resistance rule. Indeed, the correlation between the
oscillatory diagonal- and off-diagonal- resistances held true down
to nearly $\nu = 5$, see left inset of Fig. 5(e), as narrow IQHE
plateaus were manifested in $R_{xy}$, see Fig. 5(e). For $\nu \le
4$, however, $R_{xx}$ correlated better with $B[dR_{xy}/dB]$, see
right inset of Fig. 5(e), than with $- \Delta R_{xy}$, which
suggested that the resistivity/resistance rule\cite{50} might come
into play at especially low-$\nu$ here, as the system undergoes a
Type-3 $\rightarrow$ Type-1 transformation, with decreasing $\nu$.

An expanded data plot of Type-3 transport is provided in Fig.
6(a). This plot shows plateaus in $R_{xy}$ and deep minima in
$R_{xx}$ to very low-B, as quantum Hall plateaus in $R_{xy}$
follow $R_{xy} = h/ie^{2}$, to an experimental uncertainty of
$\approx 1$ percent. Although the IQHE data of Fig. 6(a) again
appear normal at first sight, the remarkable difference becomes
apparent when $\Delta R_{xx}$ and -$\Delta R_{xy}$ are plotted
\textit{vs}. $\nu$, as in Fig. 6(b). Here, we find once again an
approximate phase-shift of "$\pi$" ("Type-3") between  $\Delta
R_{xx}$ and $\Delta R_{xy}$ (Fig. 6(b)), as in Fig. 5(b) and (c),
that is distinct from the canonical ("Type-1") phase relationship
exhibited in Fig. 1(c). The resistivity/resistance rule $R_{xx}
\propto B[dR_{xy}/dB]$ seems not to be followed in this case at
such large $\nu$, as a "phase shift" and a lineshape difference
between $R_{xx}$ and $B[dR_{xy}/dB]$ becomes perceptible, see Fig.
6(c). Notably, in Fig. 6(a), the reported "Type-3" phase relation
can even be discerned by a trained eye.
\section{discussion: Numerical simulations indicating the possibility of "Type-3" and "Type-2" oscillations}
It is possible to extract some understanding from the phase
relationships observed here between the oscillatory $R_{xx}$ (or
$\Delta R_{xx}$) and $\Delta R_{xy}$. The "Type-1" orthogonal
phase relation of Fig. 1(c) can be viewed as a restatement of the
empirical resistivity/resistance rule, since the data of Fig. 1(a)
yield both Fig. 1(b) and Fig. 1(c). Theory suggests that this rule
might follow when $R_{xx}$ is only weakly dependent on the local
diagonal resistivity $\rho_{xx}$ and approximately proportional to
the magnitude of fluctuations in the off-diagonal resistivity
$\rho_{xy}$, when $\rho_{xx}$ and $\rho_{xy}$ are functions of the
position.\cite{53} Thus, according to theory, specimens following
the resistivity rule (and exhibiting "Type-1" oscillations) seem
likely to include density fluctuations.\cite{53}

For "Type-2" and "Type-3" oscillations, note that the specimens of
Figs. 2 - 6 satisfy $\omega \tau_{T} > 1$, with $\omega$ the
cyclotron frequency, and $\tau_{T}$ the transport lifetime, at $B
> 0.001$ (or $0.002$) $T$. One might semi-empirically
introduce oscillations into the diagonal conductivity,
$\sigma_{xx}$, as $\sigma_{xx} = \sigma_{xx}^{0}(1 - Acos(2\pi
E_{F}/\hbar\omega ))$.\cite{54,55,56} Here, the minus sign ensures
the proper phase, while $\sigma_{xx}^{0} = \sigma_{0}/(1+(\omega
\tau_{T})^{2})$, $\sigma_{0}$ is the $dc$ conductivity, $E_{F}$ is
the Fermi energy, and $A = 4c[(\omega \tau_{T})^{2}/(1+(\omega
\tau_{T})^{2})][X/sinh(X)]exp(- \pi/\omega \tau_{S})$, where $X =
[2 \pi^{2}k_{B}T/\hbar \omega]$, $\tau_{S}$ is single particle
lifetime, and $c$ is of order unity.\cite{54,56,57} Simulations
with $\sigma_{xx}$ as given above, and $\sigma_{xy} =
\sigma_{xy}^{0} =(\omega \tau_{T})\sigma_{xx}^{0}$, indicate
oscillations in both $R_{xx}$ and $R_{xy}$ via $\rho_{xx} =
\sigma_{xx} /(\sigma_{xx}^{2}+ \sigma_{xy}^{2})$ and $\rho_{xy} =
\sigma_{xy} /(\sigma_{xx}^{2}+ \sigma_{xy}^{2})$, and a Type-3
phase relationship, see Fig. 7(a), with $|\Delta R_{xy}| <<
|\Delta R_{xx}|$. That is, an oscillatory $\sigma_{xx}$ can also
lead to small $R_{xy}$ oscillations, with "Type-3" phase
characteristics, see Fig. 7(a).

As a next step, one might introduce an oscillatory $\sigma_{xy} =
\sigma_{xy}^{0}(1 + Gcos(2 \pi E_{F}/ \hbar \omega))$, where $G =
2c[(1+3( \omega \tau_{T})^{2})/((\omega \tau_{T})^{2}(1+(\omega
\tau_{T})^{2}))] [X/sinh(X)]exp(-\pi/\omega
\tau_{S})$.\cite{55,56} Upon inverting the tensor including
oscillatory $\sigma_{xy}$ and $\sigma_{xx}$, "Type-3" oscillations
were still obtained, as in Fig. 7(a).
\begin{figure}[t]
%h=here, t=top, b=bottom, p=separate figure page
\begin{center}
\leavevmode \epsfxsize=3.25in \epsfbox {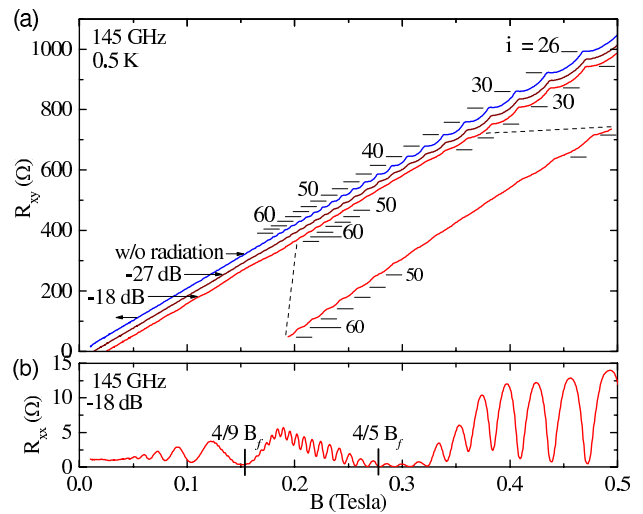}
\end{center}
\caption {Photo-excitation eliminates IQHE: (a) The dark- and
irradiated-at-$145 GHz$- off-diagonal Hall resistances $R_{xy}$ of
a GaAs/AlGaAs device at $T = 0.5 K$ have been exhibited
\textit{vs}. the magnetic field, $B$. For the sake of
presentation, the $-27 dB$ and $-18dB$ curves have been shifted
down along the ordinate with respect to the dark "w/o radiation"
curve . The index, $i$, labels the Hall plateaus. Also shown in
the figure panel is an expanded $R_{xy}$ plot of the $-18 dB$
curve for the span $34 \leq i \leq 62$, which shows that Hall
plateaus disappear under photoexcitation between roughly $36 \leq
i \leq 48$. Thus, photo-excitation eliminates these quantum Hall
effects in favor of an ordinary Hall effect. (b) This panel
exhibits the diagonal resistance $R_{xx}$ under photo-excitation
at $145 GHz$, with the photoexcitation attenuated to $-18 dB$.
Note the reduction in the the amplitude of Shubnikov-de Haas
oscillations, most noticeably in the vicinity of the $(4/5) B_{f}$
minimum of the radiation-induced magneto-resistance oscillations.
Thus, this figure shows that a radiation-induced reduction in the
amplitude of Shubnikov-de Haas oscillations in $R_{xx}$ correlates
with the radiation-induced disappearance of IQHE plateaus in
$R_{xy}$.} \label{mani05fig}
\end{figure}

Finally, the strong $B$-field $\sigma_{xy}$ follows $\sigma_{xy} =
\sigma_{xx}/\omega \tau - ne/B$ in the self-consistent Born
approximation for short range scattering potentials, when $\tau$
is the relaxation time in the $B$-field.\cite{55} Although, the
dominant scattering mechanism is long-ranged in GaAs/AlGaAs
devices, we set $\sigma_{xy} = (\sigma_{xx}^{0}/\omega \tau -
ne/B) - A(\sigma_{xx}^{0}/\omega \tau)cos(2\pi E_{F}/\hbar
\omega)$. When $\tau = \tau_{T}$, this approach again yielded
Type-3 phase relations, as in the discussion above. At this point,
we were surprised to see that the Type-3 oscillations reported
here could be so readily generated from the simulations. Next, we
examined the case $\tau < \tau_{T}$, in order to account for the
possibility that $\tau$ in a $B$-field may possibly come to
reflect $\tau_{S}$, which typically satisfies $\tau_{S} <
\tau_{T}$ for small angle scattering by long-range
scattering-potentials.\cite{57} Remarkably, a reduction in $\tau$,
which corresponds to changing the nature of the potential
landscape, converted "Type-3" (phase-shift by $\pi$) to "Type-2"
(in-phase) oscillations, see Fig. 7(a) and 7(b).
\begin{figure}[t]
%h=here, t=top, b=bottom, p=separate figure page
\begin{center}
\leavevmode \epsfxsize=3.25in \epsfbox {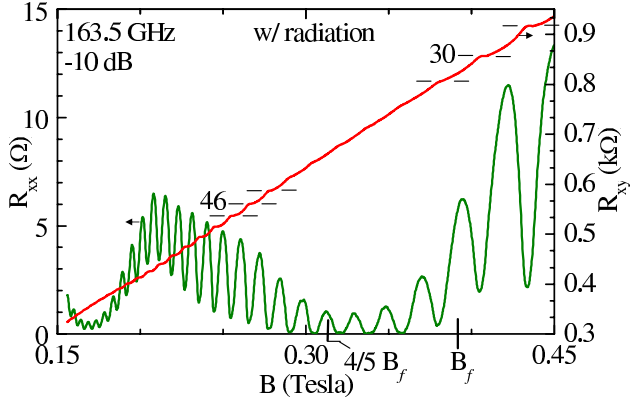}
\end{center}
\caption {Re-entrant IQHE under photo-excitation at $T = 0.5 K$:
The diagonal ($R_{xx}$) and off-diagonal ($R_{xy}$) resistances of
a photo-excited GaAs/AlGaAs device have been exhibited
\textit{vs}. the magnetic field, $B$, for $0.15 \leq B \leq 0.45
Tesla$. Here, the specimen has been photo-excited at $163.5 GHz$,
with the intensity attenuated to $-10dB$. The "slow" oscillatory
structure in $R_{xx}$ corresponds to the radiation-induced
magneto-resistance oscillations, while the "fast" structure
corresponds to the Shubnikov-de Haas oscillations. The $(4/5)
B_{f}$ minimum for the radiation-induced magneto-resistance
oscillations has been marked in the figure. Noticeably, the
amplitude of the Shubnikov-de Haas oscillations in $R_{xx}$ is
reduced in the vicinity of the minima of the radiation-induced
magneto-resistance oscillations, and this feature correlates with
a vanishing of IQHE in $R_{xy}$ over the same interval.
Specifically, in this instance, Hall plateaus vanish  between $34
\leq i \leq 42$. They reappear at higher $i$, only to disappear
once again at even higher $i$. That is, there appears to be a
radiation-induced re-entrance into IQHE. Here, the plateau index
$i$ has been marked next to the Hall plateaus. } \label{mani05fig}
\end{figure}

If density fluctuations at large length scales produce "Type-1"
characteristics,\cite{53} and "Type-2" oscillations require a
difference between $\tau_{T}$ and $\tau$ as suggested above, then
the observation of "Type-1" and "Type-2" oscillations in the same
measurement (Fig. 3(b) and (c)) seems consistent because long
length-scale potential fluctuations can produce both modest
density variations and a difference between $\tau_{T}$ and
$\tau_{S}$ (or $\tau$).\cite{57} Perhaps, with increasing $B$,
there is a crossover from "Type-2" to "Type-1" before $R_{xy}$
plateaus become manifested, and thus, IQHE is not indicated in the
"Type-2" regime, see Figs. 3 (a) and 4(a).
\begin{figure}[t]
%h=here, t=top, b=bottom, p=separate figure page
\begin{center}
\leavevmode \epsfxsize=2.5in \epsfbox {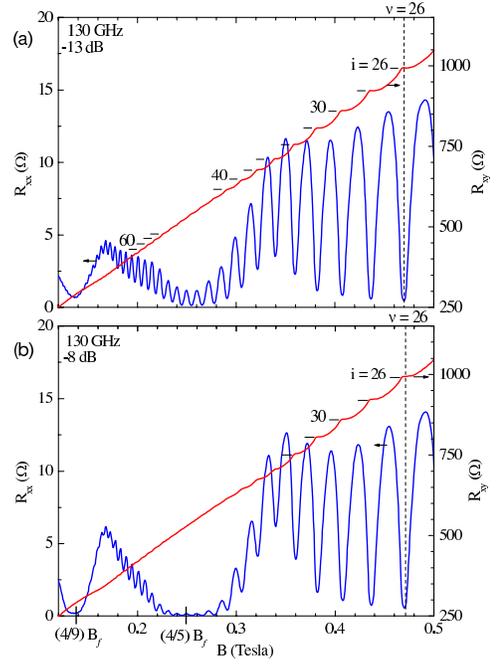}
\end{center}
\caption {Evolution of transport and re-entrant IQHE under
photo-excitation at $T = 0.5 K$: The diagonal ($R_{xx}$) and
off-diagonal ($R_{xy}$) resistances of a photo-excited GaAs/AlGaAs
device have been exhibited \textit{vs}. the magnetic field, $B$,
for $0.12 \leq B \leq 0.5 Tesla$. Here, the specimen has been
photo-excited at $130 GHz$, with the intensity attenuated to
$-13dB$ in the panel (a), and $-8dB$ in the panel (b) . Thus, a
comparison of panels (a) and (b) serves to convey the effect of an
incremental change in the radiation intensity. The bottom panel
corresponds to the greater intensity. As the radiation intensity
increases, the radiation-induced magneto-resistance oscillations
in $R_{xx}$, i.e., the "slow" oscillations, become more pronounced
and a radiation-induced zero-resistance state becomes perceptible
in $R_{xx}$ in the vicinity of $(4/5) B_{f}$. Concurrently, the
amplitude Shubnikov-de Haas oscillations in $R_{xx}$ shows even
stronger non-monotonicity \textit{vs}. $B$. Indeed, the
Shubnikov-de Haas oscillations nearly vanish in the vicinity of
$(4/5) B_{f}$ as a consequence of photo-excitation at $-8 dB$. A
comparison of the top- and bottom- panels also shows that Hall
plateaus become weaker and tend to vanish with increased
photo-excitation near the minima of the radiation-induced
magneto-resistance oscillations. The effect tracks the
disappearance of Shubnikov-de Haas oscillations in $R_{xx}$ under
the influence of photo-excitation. Indeed, some Hall plateaus that
are observable under photo-excitation at $-13 dB$ seem to have
nearly vanished under irradiation at $-8 dB$.} \label{mani05fig}
\end{figure}
\begin{figure}[t]
%h=here, t=top, b=bottom, p=separate figure page
\begin{center}
\leavevmode \epsfxsize=2.5in \epsfbox {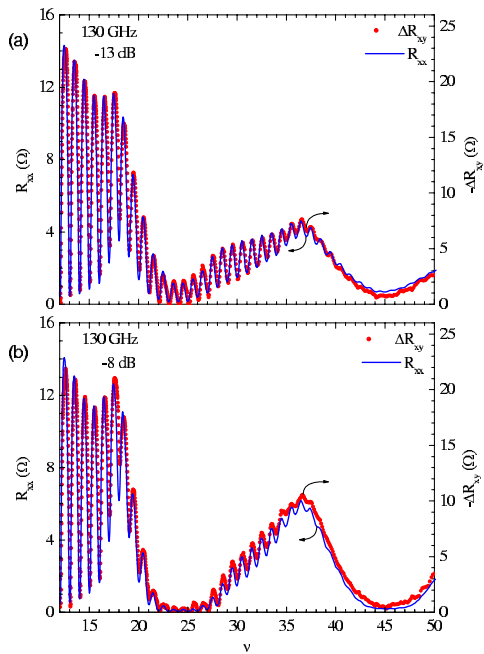}
\end{center}
\caption {"Type-3" transport under photo-excitation: This figure
presents a re-plot of the data of the previous figure, Fig. 10, as
$R_{xx}$ \textit{vs}. $\nu$ and $-\Delta R_{xy}$ \textit{vs}.
$\nu$, in order to carry out a phase comparison of the
oscillations in the diagonal and off-diagonal resistance. Here,
the oscillatory part of the Hall resistance, $\Delta R_{xy}$,
matches, within a scale factor, the diagonal resistance so far as
both the radiation-induced- and Shubnikov-de Haas- oscillations
are concerned. Indeed, the "Type-3" behavior observed in this
figure confirms that the IQHE exhibited in the previous figure,
Fig. 10, corresponds to the new IQHE, discussed in association
with Fig. 5 and Fig. 6.} \label{mani05fig}
\end{figure}

Specimens exhibiting "Type-3" oscillations and associated IQHE
suggest better homogeneity in $n$, which is confirmed by
oscillations to extremely low-$B$ (see Fig. 6(a) and (b)). The
relatively narrow plateaus at high-$B$ (see Fig. 5(e)) hint at a
reduced role for disorder-induced localization.\cite{2} In this
case, perhaps there are other mechanisms contributing to the
observed large $|\Delta R_{xy}|$, and ("Type-3") plateau
formation, in the high-$\mu$ system. It could be that such
additional mechanisms serve to create/maintain/enhance the
mobility gap or suppress backscattering in the higher Landau level
here, which assists in the realization of the "Type-3"
characteristics and IQHE to low $B$.

\section{Part 2}
\section{Transport study under microwave photo-excitation}
Previous sections indicated the observability of several possible
phase relations, denoted as "Type-1", "Type-2", and "Type-3,"
between the oscillatory Hall and diagonal resistances in the dark
GaAs/AlGaAs system. Of particular interest were the approximately
"$\pi$"-shifted "Type-3" oscillations (Figs. 5 and 6), which
exhibited quantum Hall effects that were similar to,- and yet
different from,- the canonical quantum Hall effect that goes
together with the resistivity/resistance rule and "Type-1"
oscillations characterized by an approximately "$\pi/2$"
phase-shift (Fig. 1). Numerical simulations exhibited in the last
section confirmed the possibility of these "Type-3" and "Type-2"
phase relations, although the magnitude of the Hall oscillations
obtained in the simulations always remained small compared to
$R_{xx}$ oscillations, unlike experiment. In this section, we
examine the influence of microwave photo-excitation over the same
magnetic field, filling factor, and temperature intervals, to
determine- and convey- the radiation-induced change in
Shubnikov-de Haas oscillations and IQHE in the "Type-3" high
mobility system. The MBE grown GaAs/AlGaAs single heterostructures
for these measurements were prepared by Umansky and coworkers, as
per ref. 58.

Figure 8(a) illustrates the dark- and irradiated-at-$145 GHz$-
off-diagonal Hall resistances $R_{xy}$ of a GaAs/AlGaAs device at
$T = 0.5 K$. Here, for the sake of presentation, the $-27 dB$ and
$-18dB$ Hall curves have been down-shifted along the ordinate with
respect to the dark "w/o radiation" Hall curve, and the Hall
plateaus have been labeled with the index, $i$. Fig. 8(a) shows
that the $R_{xy}$ curve is influenced by the radiation in two
ways: First, there develops the "slow" radiation-induced
oscillations in the Hall resistance, which are superimposed on the
overall linear increase in $R_{xy}$ with $B$. These
radiation-induced oscillations in $R_{xy}$ are especially evident
in the $-18 dB$ curve, and they are similar to what we have
reported earlier.\cite{9}. Second, radiation also appears to
narrow the width of the Hall plateaus more readily over some range
of filling factors than others. Again, this feature is most
evident in the $-18 dB$ curve. Indeed, in the expanded $R_{xy}$
plot of the $-18 dB$ curve in Fig. 8(a) for the span $34 \leq i
\leq 62$, the Hall plateaus tend to vanish under photoexcitation
between roughly $36 \leq i \leq 48$, only to reappear at even
higher $i$. Thus, this figure suggests that photo-excitation
exchanges the integral quantum Hall effects with an ordinary Hall
effect over some range of filling factors. The bottom panel of
Fig. 8 exhibits the diagonal resistance $R_{xx}$ under
photo-excitation at $f$ = $145 GHz$, with the photoexcitation
attenuated to $-18 dB$. This panel exhibits strong
radiation-induced magneto-resistance oscillations in $R_{xx}$, and
a reduction in the amplitude of Shubnikov-de Haas oscillations,
most noticeably in the vicinity of the $(4/5) B_{f}$ minimum,
where $B_{f} = 2 \pi m^{*} \textit{f}/e$.  A comparison of the top
and bottom panels of Fig. 8 indicates that the radiation-induced
vanishing of Hall plateaus in $R_{xy}$ correlates with this
radiation-induced non-monotonicity in the amplitude of
Shubnikov-de Haas oscillations in $R_{xx}$.

Figure 9 exhibits similar characteristics at a radiation frequency
$f = 163.5 GHz$. Here, a simultaneous measurement of $R_{xx}$ and
$R_{xy}$, followed by a plot of the two curves on the same graph,
helps to convey the intimate relation between the
radiation-induced non-monotonic variation in the amplitude of the
Shubnikov-de Haas $R_{xx}$ oscillations and the non-monotonic
variation in the width of the Hall plateaus in $R_{xy}$. This plot
suggests that as the amplitude of Shubnikov-de Haas oscillations
in $R_{xx}$ is reduced by photo-excitation, for example, in the
vicinity of the $(4/5) B_{f}$ radiation-induced resistance
minimum, the Hall plateaus tend to narrow and vanish, although the
plateaus reappear once again at either end of the $(4/5) B_{f}$
minimum, when the Shubnikov-de Haas oscillations in $R_{xx}$ grow
stronger.

Figure 10 helps to convey the evolution of these transport
characteristics with the photo-excitation intensity at $f= 130
GHz$. Hence, Fig. 10(a) shows $R_{xx}$ and $R_{xy}$ \textit{vs}.
$B$ with the radiation attenuated to $-13dB$, while Fig. 10(b)
shows the same with the radiation attenuated to $-8 dB$. Here,
Fig. 10(b) corresponds to the more intense photo-excitation.
Although the two panels look similar, it is apparent that
increasing the intensity from $-13 db$ to $-8 dB$ increases the
magnitude of the "slow" radiation-induced magnetoresistance
oscillations characterized by relatively broad minima in the
vicinity of $(4/5) B_{f}$ and $(4/9) B_{f}$. Concurrently, the
amplitude of Shubnikov-de Haas oscillations in $R_{xx}$ becomes
smaller in the vicinity of, for example, the $(4/5) B_{f}$
minimum. This weakening of SdH oscillations goes together again
with the progressive disappearance of IQHE. It is worth pointing
out that IQHE were easily observable down to approximately $0.2
Tesla$ at this temperature, $T=0.5K$, in the absence of
photo-excitation, as in Fig. 6. Fig. 10 shows, however, that under
photo-excitation, although some IQHE are still observable in the
vicinity of $0.2 Tesla$, they disappear in the vicinity of $0.25
Tesla$ as a consequence of the photo-excitation, only to reappear
at an even higher $B$.

The data of Figs. 8 - 10 help to illustrate that microwave
photo-excitation not only produces radiation-induced
magneto-resistance oscillations, but it also influences the
amplitude of Shubnikov-de Haas oscillations, producing a
non-monotonic variation in the amplitude of these oscillations
\textit{vs}. $B$ or $\nu$. Further, the data show that this
non-monotonic variation in the amplitude of the Shubnikov-de Haas
oscillations correlates with a non-monotonic variation in the
width of the IQHE plateaus \textit{vs}. $B$ or $\nu$. As a
vanishing Hall plateau width signals the disappearance of IQHE, it
appears that photo-excitation helps to replace the IQHE with an
ordinary Hall effect over the minima of the radiation-induced
magneto-resistance oscillations.

In order to identify the mechanism responsible for the
disappearance of the Hall plateaus under photo-excitation, it
seems necessary to establish first the nature of IQHE at these
high filling factors per the analysis used earlier for examining
transport in the dark specimens. Hence, in Fig. 11, we re-plot the
the data of Fig. 10 as $R_{xx}$ \textit{vs}. $\nu$ and $\Delta
R_{xy}$ \textit{vs}. $\nu$ . The plots of Fig. 11 are generally
similar to the plots of Fig. 1(c), Fig. 3(b) and (c), Fig. 4(b),
Fig. 5(b) and (c), and Fig. 6(b), with the difference that,
instead of $\Delta R_{xx}$, it is $R_{xx}$ that is directly
compared with -$\Delta R_{xy}$, in the plot overlays. The reason
for this difference is that we wish to compare, at the same time,
the relative phases of the "fast" "SdH oscillations," as well as
the "slow" radiation-induced resistance oscillations, in $R_{xx}$
and $R_{xy}$.

In previous work,\cite{9} we have shown that so far as the
radiation-induced oscillations in $R_{xx}$ and $R_{xy}$ are
concerned, -$\Delta R_{xy}$ $\propto$ $R_{xx}$. By setting up this
same relation in the plots of Fig. 11, we notice that both the
radiation-induced and "SdH" oscillations in the diagonal and
off-diagonal resistances match up over nearly the entire range of
exhibited filling factors, $12 \leq \nu \leq 50$. Thus, Fig. 11
helps to realize three conclusions: (a) There is an approximate
"$\pi$"-phase shift between the Shubnikov-de Haas oscillations of
the diagonal- and off-diagonal Hall resistances in these
photo-excited data, as for the "Type-3" resistance oscillations
sketched in Fig. 2(c). (b) Radiation-induced magneto-resistance
oscillations of the diagonal and off-diagonal resistances follow
the ("Type-3") relation -$\Delta R_{xy}$ $\propto$ $R_{xx}$, as
previously reported.\cite{9} (c) The IQHE observed in this
photo-excited study exhibit the "Type-3" characteristic,
\textit{cf}. Fig. 10 and 11, once again, as in the dark study,
\textit{cf}. Fig. 5 and 6. Thus, the observed IQHE seem to belong
once again to this new class of IQHE.

These strong correlations between the $R_{xx}$ and $-\Delta
R_{xy}$ characteristics per Fig. 11 further suggest that the
mechanism that is responsible for the radiation-induced modulation
in the amplitude of Shubnikov-de Haas oscillations in $R_{xx}$ is
also the mechanism that produces the modulation in the width of
the IQHE. This topic will be examined in greater detail
elsewhere.\cite{59}

Finally, the observation of vanishing diagonal resistance and an
ordinary Hall effect under photo-excitation, over a range of
filling factors where vanishing resistance is associated with
quantum Hall effect in the dark specimen, suggests that the
disordered 2D system attains the appearance of an ideal,
disorder-free, two-dimensional electron system, under
photo-excitation.\cite{2}

\section{Conclusion}

An experimental study of the high mobility GaAs/AlGaAs system in
the dark at large-$\nu$ indicates three distinct phase relations
between the oscillatory Hall and diagonal resistances, which have
been labeled as "Type-1", "Type-2", and "Type-3" oscillations.
"Type-1" corresponds to the canonical quantum Hall situation at
large filling factors. "Type-2" and "Type-3" oscillations can be
reproduced in simple transport simulations, and such oscillations
exhibit systematic deviations from the resistivity/resistance
rule. Surprisingly, IQHE appears manifested in high mobility
specimens in the case of "Type-3" oscillations, which are
characterized by approximately "anti-phase" Hall- and diagonal-
resistance oscillations. Based on the differences at large $\nu$
between IQHE in the "Type-3" case and the canonical IQHE, so far
as the phase relations, and consistency with the
resistivity/resistance rule are concerned, we have reasoned the
"Type-3" case corresponds to a new class of IQHE.

We have also examined the influence of photo-excitation at
microwave frequencies on the Hall and diagonal resistances in the
high mobility specimen at large filling factors. We have observed
that photo-excitation serves to produce a non-monotonic variation
with $B$ or $\nu$ in the amplitude of Shubnikov-de Haas
oscillations, and concurrently modulates the width of the IQHE
plateaus, leading, remarkably, to a vanishing of IQHE at the
minima of the radiation-induced magneto-resistance oscillations.
Strikingly, the "Type-3" phase relation is observed for both the
"SdH" oscillations- and the radiation-induced magneto-resistance
oscillations- in $R_{xx}$ and $R_{xy}$, see Fig. 11. The results
suggest that the mechanism that produces the modulation in the
amplitude of SdH oscillations in $R_{xx}$ is also be responsible
for plateau narrowing and IQHE quenching at the minima of the
radiation-induced magneto-resistance oscillations.
\section{Acknowledgement}
R.G.M. is supported by D. Woolard and the Army Research Office
under W911NF-07-01-0158.

\begin{figure}[t]
%h=here, t=top, b=bottom, p=separate figure page
\begin{center}
\leavevmode \epsfxsize=3in \epsfbox {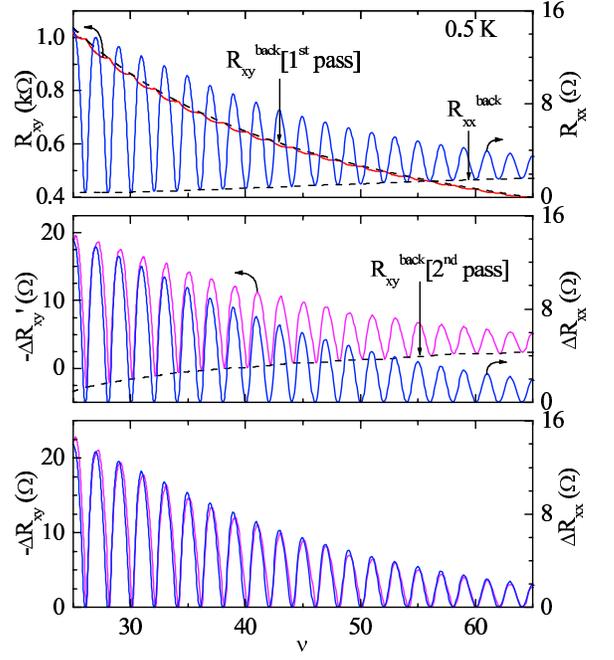}
\end{center}
\caption {Background subtraction procedure: (Top) The diagonal and
off-diagonal resistances $R_{xx}$ and $R_{xy}$ of a GaAs/AlGaAs
device have been exhibited \textit{vs}. the filling factor $\nu$
along with $R_{xx}^{back}$ and $R_{xy}^{back}$[1st pass], the
background Hall resistance at the first pass. $R_{xy}^{back}$[1st
pass] is obtained through a linear fit of $R_{xy}$ \textit{vs}.
$B$. (Center) This panel exhibits $\Delta R_{xx} = R_{xx} -
R_{xx}^{back}$ and $-\Delta R_{xy}^{'} =
-(R_{xy}-R_{xy}^{back})$[1st pass]. (Bottom) This panel exhibits
$\Delta R_{xx}$ and $-\Delta R_{xy} = -(\Delta R_{xy}^{'}
-R_{xy}^{back})$[2nd pass].} \label{mani05fig}
\end{figure}
%\appendix*
\section{Appendix: Background subtraction and the extraction of the
oscillatory resistances} The purpose of this appendix is to
exhibit the background subtraction procedure that has been used to
extract the oscillatory resistances and generate the resistance
oscillations overlays in the figures, for phase comparison. For
the diagonal resistance, $R_{xx}^{back}$, when utilized, typically
followed either the midpoints (e.g. Fig. 1(a)) or the minima (e.g.
Fig. 4(a)) of the $R_{xx}$ oscillations, see also Fig. 12(Top).
The resulting $R_{xx}^{back}$ is then removed from $R_{xx}$ to
obtain $\Delta R_{xx}$ as shown in Fig. 12(center). Background
subtraction for the off-diagonal Hall resistance involved a two
pass process, since $|R_{xy}^{back}| >> |\Delta R_{xy}|$. Here,
the first pass identified $\approx 99\%$ of $R_{xy}^{back}$
through a linear-fit of $R_{xy}$ \textit{vs}. $B$  which is shown
in Fig. 12(Top), while a spline fit in the second pass, see Fig.
12 (center), then accounted for the $\approx 1\%$ residual term.
At the second pass, $R_{xy}^{back}$ was chosen as for $R_{xx}$,
see above, to make possible resistance oscillations overlays of
the type shown in Fig. 12(bottom). In all cases, $R_{xx}^{back}$
and $R_{xy}^{back}$ varied "slowly" in comparison to the
oscillatory part of the resistances.

\end{document}